# Development of Supersonic Plasma Flows by use of a Magnetic Nozzle and an ICRF Heating

M. Inutake, A. Ando, K. Hattori, H. Tobari, Y.Hosokawa, R. Sato, M. Hatanaka and K. Harata

*Depertment of Electrical Engineering, Tohoku University, Aoba05, sendai 980-8579, Japan;*
*inutake@ecei.tohoku.ac.jp*

## ABSTRACT

*A high-beta, supersonic plasma flow plays a crucial role in MHD phenomena in space and fusion plasmas. There are a few experimental researches on production and control of a fast flowing plasma in spite of a growing significance in the magnetized-plasma flow dynamics.*

*A magneto-plasma-dynamic arcjet (MPDA) is one of promising devices to produce a supersonic plasma flow and has been utilized as an electric propulsion device with a higher specific impulse and a relatively larger thrust. We have improved the performance of an MPDA to produce a quasi-steady plasma flow with a transonic and supersonic Mach number in a highly-ionized state.*

*There are two methods in order to control an ion-acoustic Mach number of the plasma flow exhausted from an MPDA: one is to use a magnetic Laval nozzle to convert a thermal energy to a flow energy and the other is a combined system of an ion heating and a divergent magnetic nozzle. The former is an analogous method to a compressible air flow and the latter is the method proposed in an advanced thruster for a manned interplanetary space mission.*

*We have clarified the plasma flow characteristics in various shapes of a magnetic field configuration. It was demonstrated that the Mach number of the plasma flow could increase up to almost 3 in a divergent magnetic nozzle field. This paper reports recent results on the flow field improvements: one is on a magnetic-Laval-nozzle effects observed at the muzzle region of the MPDA, and the other is on ICRF (ion-cyclotron-range of frequency) heating of a supersonic plasma by use of a helical antenna.*

## I. INTRODUCTION

Electric propulsion (EP) is one of the most promising space propulsions due to its high specific impulse $I_{sp}$. $I_{sp}$ is defined by a ratio between thrust and the flow rate of propellant weight, representing the exhaust velocity divided by the gravity acceleration. This system utilizes the electric energy to ionize a propellant gas. As the ionized propellant is exhausted downstream of the thruster with a velocity much higher than that in a chemical rocket, EP system enables a long-term space mission with a less consumption of the propellant. Recently, they are utilized not only for an attitude or position control for various satellites but also for a main engine in an asteroid-sample-return project Muses-C of ISAS, Japan launched in May, 2003 [1].

For a manned interplanetary space thruster, both a higher specific impulse and a larger thrust are required. Until the realization of a fusion-plasma thruster, an MPD thruster driven by a fission reactor is one of the promising candidates for a manned Mars spacecraft. The MPDA plasma is accelerated axially by a self-induced $j \times B$ force [2]. Thrust performance of an MPDA is expected to improve by applying an external magnetic nozzle instead of a solid nozzle [3],[4]. In order to investigate effects of the external magnetic field on the flow characteristics, a plasma flow produced by an MPDA in the HITOP device, Tohoku University is studied in detail by use of a spectrometer and Mach probes [5]–[8].

As an another advanced thruster, the Variable Specific Impulse Magneto-plasma Rocket (VASIMR) is proposed by NASA [9]. The VASIMR is a high power, radio frequency-driven magneto-plasma rocket. The plasma is produced by a helicon wave and a kinetic energy is added into ions by RF (radio frequency) heating and a magnetic nozzle transforms the resultant thermal energy into an axial flow energy. Ion heating condition in a fast flowing plasma would be different from that in a confined, stationary fusion plasma on several points such as the short transit time for ions to pass only once through a heating region and the variation of the resonance frequency due to the Doppler shift effect.

In this paper are presented two kinds of basic experiments to improve the acceleration performance of the MPDA plasma. One is the flow characteristics in a magnetic Laval nozzle and the other is ICRF (ion-cyclotron-range of frequency) ion heating of a fast flowing plasma ejected from the MPDA in the divergent magnetic nozzle.

## II. Experimental Apparatus and Diagnostics

Experiments are performed in the HITOP (HIgh density TOhoku Plasma) device of Tohoku University. The device consists of a large cylindrical vacuum chamber (diameter $D = 0.8$ m, length $L = 3.3$ m) with external magnetic coils as shown in Fig.1. Various types of magnetic field configuration can be formed by adjusting each coil current. A high-power, quasi-steady (1ms) MPDA has a coaxial structure with a center tungsten-rod cathode (10 mm in diameter) and an an-nular molybdenum anode (30 mm in inner diameter). A fast-acting gas valve can inject helium or argon gas quasi-steadily for 2-3 ms with a rise/fall time of 0.2 ms. Arc discharge is initiated when the gas flow rate becomes steady state. Quasi-steady (1 ms) discharge current $I_d$ is supplied by a PFN (pulse forming network) power supply with the maximum $I_d$ of 10 kA and a typical discharge voltage of 200 V.

Several diagnostic instruments are installed on the HITOP device. Flow velocity $U$ and ion temperature $T_i$ in the region near the MPDA are measured by the Doppler shift and broadening of HeII line spectra ($\lambda$ = 468.58 nm) and by Mach probes in the region far downstream of the MPDA.

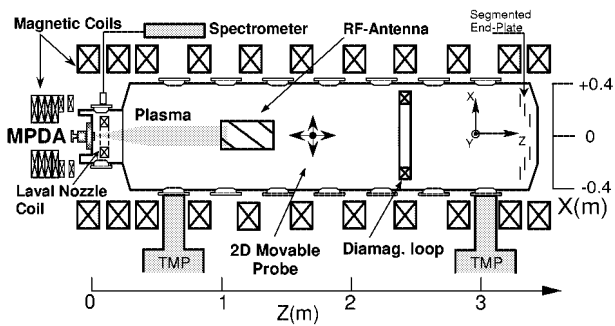

Figure 1: Schematic of the HITOP device.

The ion acoustic Mach number $M_i$ is defined as a ratio of the plasma flow velocity $U_z$ to the sound velocity $C_s$ and calculated as follows,

$$M_i = \frac{U_z}{C_s} = \frac{U_z}{\sqrt{k_B(\gamma_e T_e + \gamma_i T_i)/m_i}}, \quad (1)$$

where $\gamma_i$ and $\gamma_e$ are specific heat ratio of ion and electron, respectively. The Mach probe has two plane surfaces, one of which faces upstream of the flow, called a parallel tip, and the other faces perpendicularly to the axial flow, called a perpendicular tip, hereinafter. Effect of the magnetic field is negligible in the ion saturation current, because the ion Larmor radius is much larger than the size of the probe tip in the present experiments. The ion Mach number can be derived from a ratio of the two ion saturation current densities $J_{\text{para}}$ and $J_{\text{perp}}$, which are collected in the parallel and perpendicular tips, respectively.

$$M_i = \kappa \frac{J_{\text{para}}}{J_{\text{perp}}} \quad (2)$$

The above relation is calibrated with the spectroscopic measurements in specially-arrangement experiments [5].

### III. Experimental Results

#### III.A Improvement of Flow Characteristics by use of a Magnetic Laval Nozzle

In case of a uniform magnetic channel, it is found that both axial and rotational flow velocities of the MPDA plasma increase linearly with an increasing discharge current and at the same time, ion temperature increases more steeply. This results in the limitation of the ion Mach number less than unity in the vicinity of the muzzle [6].

From multi-channel magnetic probe measurements, it is found that the diamagnetic effect is so strong that the externally-applied uniform field is modified to an effectively converging nozzle [7]. According to a conventional gas dynamics, it is expected a supersonic plasma injected into the effectively-converging nozzle would be decelerated or suddenly jump into a subsonic flow through a shock wave. On the other hand, a subsonic plasma injected would be accelerated to a sonic flow in the converging nozzle. This could be the reason why the Mach number is always around unity in the downstream region in case of a uniform field. When the field is divergent in the downstream region, a supersonic flow has been obtained as shown in Fig.2 (region(I)). When the supersonic flow is intentionally introduced in a mirror field (effective Laval nozzle), a standing shock-wave appears in front of the Laval nozzle (region (II)) and the subsonic flow behind the shock wave is re-accelerated to a supersonic flow with $M_i$ of about 2 (region (III)) [8].

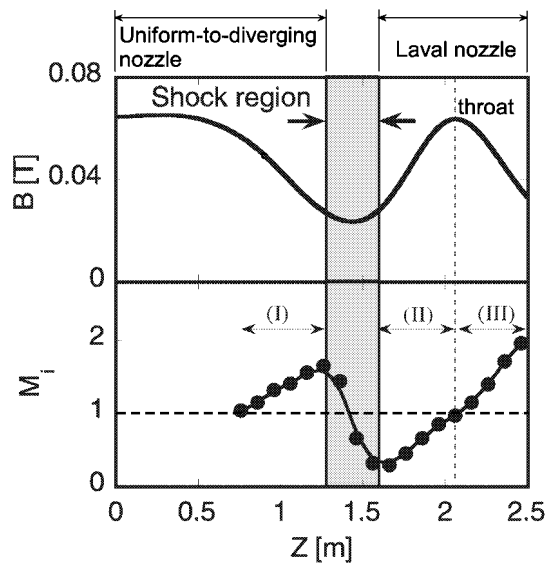

Figure 2: Formation of a standing shock wave in front of a long magnetic Laval nozzle and re-acceleration of the post-shock subsonic flow to a supersonic one through the Laval nozzle.

In order to improve the MPD thruster performance by converting a thermal energy to a flow energy, a short magnetic Laval nozzle is installed near the MPDA muzzle as shown in Fig.3. In Fig.4 are shown flow characteristics with and without the Laval nozzle. It is clear that the flow velocity and ion temperature downstream of the nozzle are higher and lower than those without the nozzle, respectively. On the other hand, variation of the flow parameters upstream of the nozzle is reversed. This shows that the subsonic flow upstream of the nozzle perceives the existence of the nozzle and self-adjusts so as to satisfy the sonic condition at the nozzle throat.

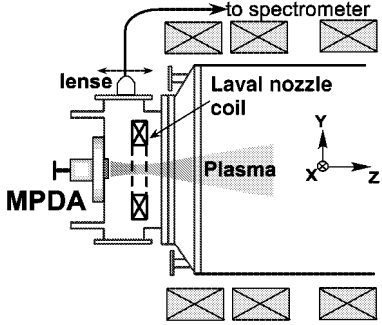

Figure 3: A short magnetic Laval nozzle installed near the MPDA muzzle.

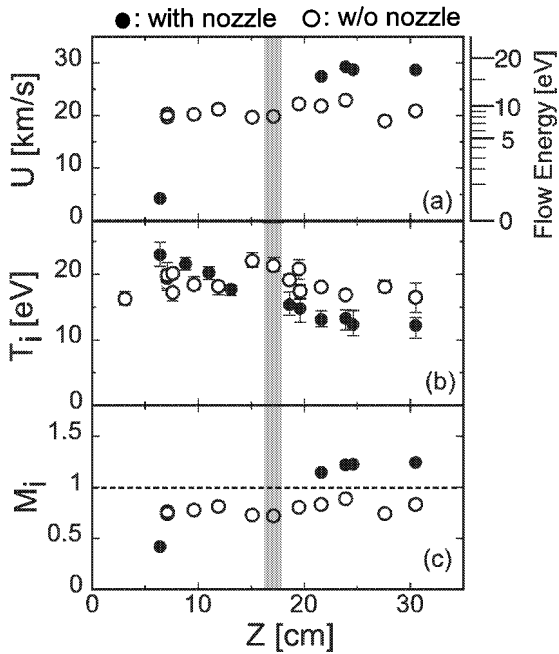

Figure 4: Flow characteristics with (solid circles) and without (open circles) the Laval nozzle. A subsonic flow is converted into a supersonic one by passing through the magnetic Laval nozzle.

It is also confirmed that total energy of the flow energy and the thermal energy in the cases with and without the Laval nozzle agrees with each other within an experimental error. The measured flow velocity, ion temperature and Mach number variation through the Laval nozzle are compared with that predicted by a 1-D isentropic flow model as shown in Fig.5. These are in a qualitative agreement with each other. Quantitatively, the Mach number measured downstream of the nozzle throat is lower than the predicted one. This would be due to an under-expansion expected from the observation that the cross section of the plasma flow downstream of the throat does not coincide with that of the vacuum magnetic channel. This effect could be reduced by increasing the magnetic field strength of the nozzle and/or by increasing the characteristic scale length of the magnetic nozzle. It is noted that an optimum converging ratio of the Laval nozzle should be selected according to the Mach number of the plasma flow ejected from the MPDA.

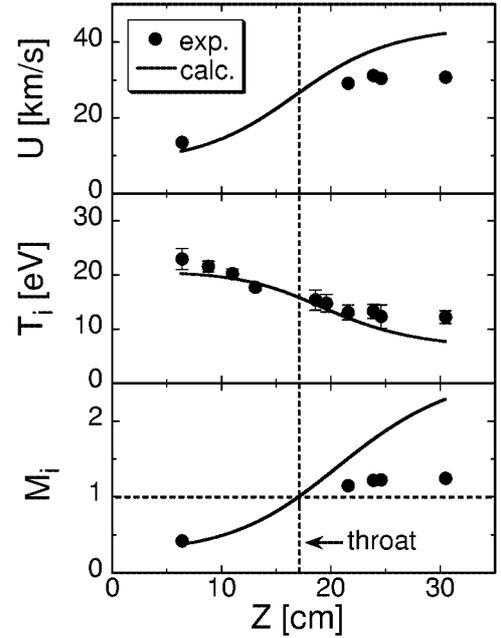

Figure 5: Comparison of a magnetic Laval Flow with the 1-D isentropic flow model with the specific heat ratio $\gamma_i$ of 5/3.

### III.B Improvement of Flow Characteristics by use of RF Heating

Ion heating of a fast-flowing plasma by a radio-frequency (RF) wave is one of attractive techniques for an advanced space propulsion. Combined with a magnetic-nozzle acceleration, the plasma flow may be accelerated efficiently by RF heating. In previous work [10], various types of antenna such as a Rogowski-type and a loop-type antennas with an azimuthal mode number $m = 0$, and a double loop antennas with $m = \pm 1$ (Nagoya type -III) and $m = \pm 2$ have been used to excite inductively waves in the ion cyclotron range of frequency. In the present work, we performed ICRF (ion cyclotron range of frequency) heating experiments by using a helical antenna as shown in Fig.6.

In order to identify propagating RF waves, wave amplitudes and phase shifts are measured by magnetic probes in the downstream region. Wavelength is obtained from the phase difference between two magnetic probe signals set at $Z = 1.03$ m and 1.43 m. Dispersion relations of the waves excited by the two types of antenna are measured in a uniform magnetic field configuration. The right helical antenna tends to excite RF wave propagating downstream with a non-axisymmetric azimuthal mode of $m = -1$, which corresponds to the slow wave, a shear Alfvén wave, in the low frequency region. The left helical antenna, on the other hand, tends to excite RF wave propagating downstream with $m = +1$, which corresponds to the fast wave, a compressional Alfvén wave [11].

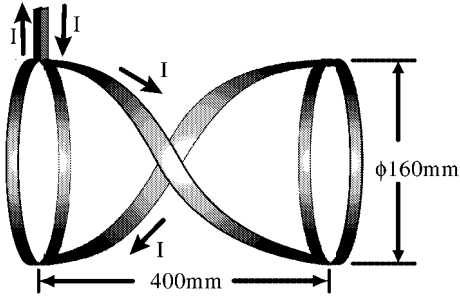

Figure 6: Schematic of the right helical antenna.

Figure 7 shows dispersion relations observed in an argon plasma with a uniform magnetic field of 0.79 kG and the ion cyclotron resonance frequency ($f_{ci} = \omega_{ci}/2\pi$) of 30 kHz in an argon plasma. The dispersion relations of shear and compressional Alfvén waves are calculated by taking account of the Doppler effect due to a plasma flow with an Alfvén Mach number $M_A$ of 0.13. These are plotted in the figure. RF waves are excited by the right or left helical antenna and the dispersion relations are obtained by varying the RF frequency. The wave dispersion excited by the right helical antenna agrees well with that of the shear Alfvén wave in the region of $\omega/\omega_{ci} < 1.5$ and seems to agree with that of the compressional one in the region of $\omega/\omega_{ci} > 2$. The wave excited by the left helical antenna agrees well with that of the compressional one in all ranges of frequency.

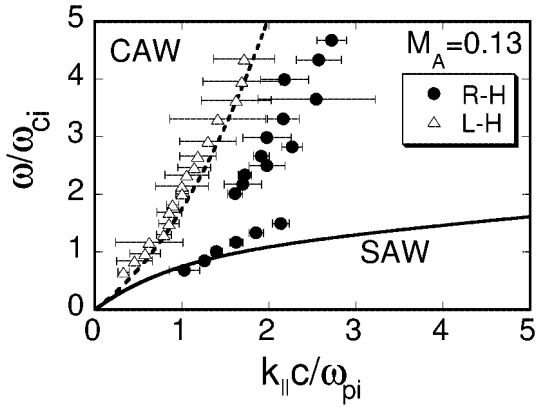

Figure 7: Dispersion relations of the propagating wave in an argon plasma. RF waves are excited by right helical (R-H, ●) and left helical (L-H, △) antennas. $B_z$ =0.79 kG(uniform). Solid and dashed lines are calculated dispersion relations of the shear and compressional Alfvén waves, respectively, with the Alfvén Mach number $M_A$= 0.13.

Figure 8 shows time evolutions of the discharge current $I_d$, the antenna current $I_{RF}$ and diamagnetic coil signal $W_\perp$. The diamagnetic coil signal apparently increases during the RF excitation period. To confirm ion heating we measured ion and electron temperatures and a spatial profile of plasma density. The ion and electron temperature $T_i$ increases from 3.9eV to 6.3eV and also electron temperature $T_e$ from 1eV to 1.5eV. The electron density is $1 \times 10^{19} \mathrm{m}^{-3}$ and only slightly decreases during the RF excitation period. The measurement time is just after the RF pulse-off time to eliminate the RF-oscillating-field effect on the probe measurement. The increment of $W_\perp$ quantitatively agrees with the increment of $T_i$ and $T_e$.

Dependences of the heating efficiency on the magnetic field strength $B_D$ in the downstream region are compared in several magnetic configurations shown in Fig.9. Ratios of $\Delta W_\perp$ to $I_{RF}$ gradually increases as $\omega/\omega_{ci}$ increases. No difference caused by the field configurations and no clear indication of the cyclotron resonance of thermal ions are observed. The reason is that the ion-ion collision frequency $\nu_{ii}$ is larger than the ion cyclotron frequency $f_{ci}$ under the experimental conditions and the waves are damped not by the cyclotron resonance but by the collisional damping. We measured the ratio of $\Delta W_\perp$ to the magnetic field amplitude $|\tilde{B}|$ of the propagating wave under several conditions of the plasma density. The ratio increases almost linearly with respect to the collision frequency. This result also indicates that the waves are damped mainly by ion-ion collisions. This damping mechanism would play an effective role in an MPD thruster operation with a high density plasma and a lower magnetic field.

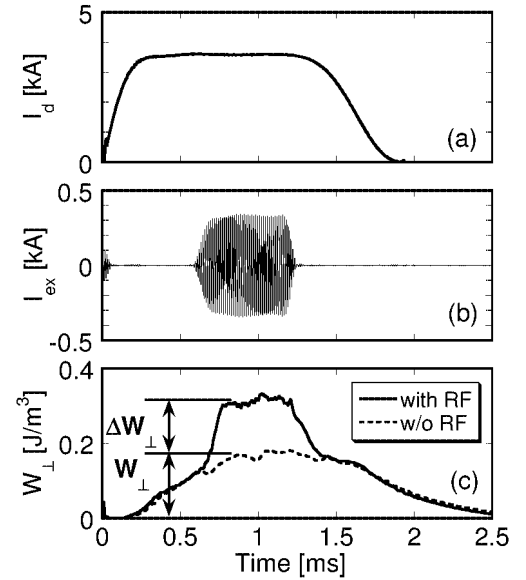

Figure 8: Temporal evolutions of (a)$I_d$, (b)$I_{RF}$ and (c)$W_\perp$. Helium plasma. $B_z = 0.7$ kG (uniform). $f_{RF}$ = 80 kHz ($\omega/\omega_{ci} = 0.3$).

In order to clarify the cyclotron resonance, we reduced the plasma density to vary the ratio of $f_{ci}$ to $\nu_{ii}$ as shown in Fig.10. Figure 11 shows dependence of the ratio $\Delta W_\perp/W_\perp$ on the magnetic field $B_D$ in the downstream region for three different plasma densities.

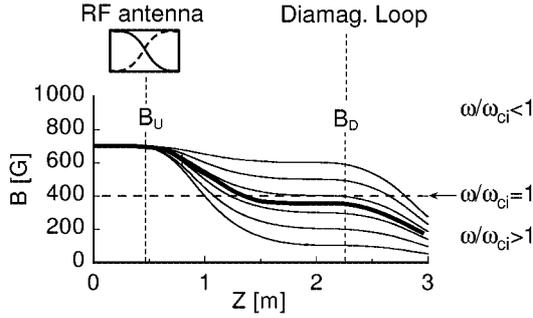

Figure 9: Axial profiles of the magnetic beach configuration with various types of the diverging nozzle ratio. The wave excited in the $B_U$ where $\omega/\omega_{ci} < 1$ propagates downstream into the $B_D$ region.

The magnetic configuration is of a magnetic-beach type with a fixed $B_U$ of 0.7 kG at the antenna position and a variable $B_D$ at the diamagnetic coil position. In Fig.11, a clear indication of the ion cyclotron resonance is observed when $f_{ci}$ is several times larger than $\nu_{ii}$. It is also observed that the peak position is shifted to a lower $B_D$, which means that the resonance occurs at $\omega/\omega_{ci}$ larger than unity. It indicates clearly the Doppler effect due to the fast plasma flow.

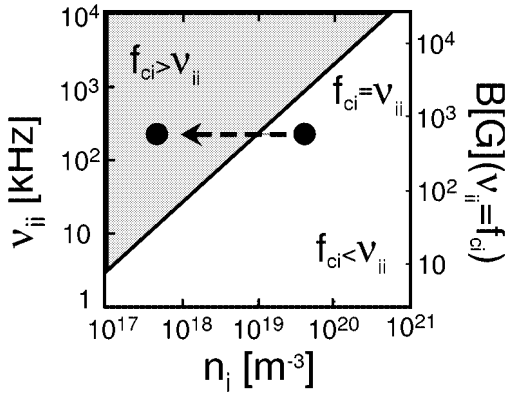

Figure 10: Dependence of the heating efficiency on ion-ion collision frequency $\nu_{ii}$. Argon plasma. $B_z$ =0.7 kG(uniform). $f_{RF}$=80 kHz.

## IV. Summary

Performance of an MPDA thruster is improved by use of a magnetic Laval nozzle and by also an RF heating. A subsonic flow near the MPDA is converted to a supersonic flow through the conversion of a thermal energy to a flow energy. The plasma flow self-adjusts so as to satisfy the sonic condition at the throat of the magnetic nozzle. An RF wave heating of a fast flowing plasma is performed in various diverging magnetic nozzles. Both shear and compressional Alfvén waves are excited and identified by comparing with theoretical curves which take into account the Doppler effect of a fast flow. A fast flowing plasma is successfully heated by use of a helical ICRF antenna with an azimuthal mode $m = \pm 1$ in the diverging magnetic nozzle. A collisional damping is dominated in the RF heating of a higher density plasma. An ion cyclotron resonance heating in a magnetic beach configuration is clearly observed for the first time under a low-collisionality-plasma condition.

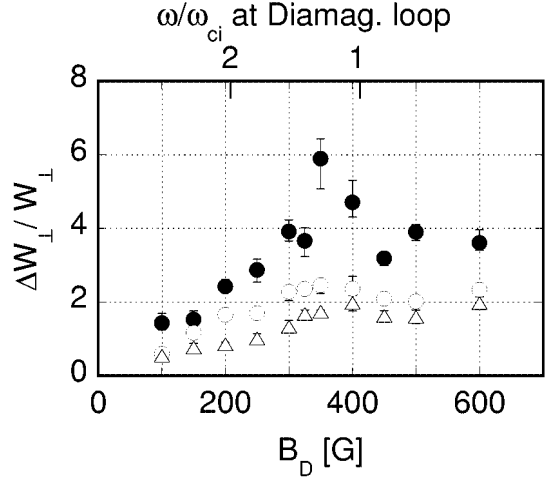

Figure 11: Incremental ratio of diamagnetic signal $\Delta W_\perp/W_\perp$ as a function of the magnetic field $B_D$ in the downstream region of a magnetic beach-field configuration. Helium plasma. $B_U$=0.7 kG. $f_{RF}$=160 kHz. (●) $f_{RF}/\nu_{ii}$=8.4 ($n = 0.52 \times 10^{18}$m$^{-3}$), (○) $f_{RF}/\nu_{ii}$=2.4 ($n = 1.9 \times 10^{18}$m$^{-3}$), (△) $f_{RF}/\nu_{ii}$=1.6 (n=2.7 $\times 10^{18}$m$^{-3}$).

## ACKNOWLEDGMENTS

This work was supported in part by a Grant-in-Aid for Scientific Research from Japan Society for the Promotion of Science. Part of this work was carried out under the Cooperative Research Project Program of the Research Institute of Electrical Communication, Tohoku University.